\documentclass[10pt]{iopart}

\usepackage{graphicx}	
\usepackage[comma]{natbib}
\usepackage{adjustbox}
\usepackage{changepage}
\usepackage[switch]{lineno} 
\usepackage{csquotes}

\usepackage{fancyhdr}

\usepackage{eso-pic}

\AddToShipoutPictureBG*{%
  \AtPageUpperLeft{%
    \hspace{0.55\paperwidth}%
    \raisebox{-5.5\baselineskip}{%
      \makebox[0pt][l]{\textnormal{Publications of the Astronomical Society of the Pacific, 133:014501}}
}}}%

\begin{document}

\title[Searching for TNOs in DES with Machine Learning]{Machine Learning for Searching the Dark Energy Survey for Trans-Neptunian Objects}

\author{B.~Henghes$^{1}$,
O.~Lahav$^{1}$,
D.~W.~Gerdes$^{2,3}$,
H.~W.~Lin$^{3}$,
R.~Morgan$^{4}$,
T.~M.~C.~Abbott$^{5}$,
M.~Aguena$^{6,7}$,
S.~Allam$^{8}$,
J.~Annis$^{8}$,
S.~Avila$^{9}$,
E.~Bertin$^{10,11}$,
D.~Brooks$^{1}$,
D.~L.~Burke$^{12,13}$,
A.~Carnero~Rosell$^{14,15}$,
M.~Carrasco~Kind$^{16,17}$,
J.~Carretero$^{18}$,
C.~Conselice$^{19,20}$,
M.~Costanzi$^{21,22}$,
L.~N.~da Costa$^{7,23}$,
J.~De~Vicente$^{24}$,
S.~Desai$^{25}$,
H.~T.~Diehl$^{8}$,
P.~Doel$^{1}$,
S.~Everett$^{26}$,
I.~Ferrero$^{27}$,
J.~Frieman$^{8,28}$,
J.~Garc\'ia-Bellido$^{9}$,
E.~Gaztanaga$^{29,30}$,
D.~Gruen$^{31,12,13}$,
R.~A.~Gruendl$^{16,17}$,
J.~Gschwend$^{7,23}$,
G.~Gutierrez$^{8}$,
W.~G.~Hartley$^{32,1,33}$,
S.~R.~Hinton$^{34}$,
K.~Honscheid$^{35,36}$,
B.~Hoyle$^{37,38,39}$,
D.~J.~James$^{40}$,
K.~Kuehn$^{41,42}$,
N.~Kuropatkin$^{8}$,
J.~L.~Marshall$^{43}$,
P.~Melchior$^{44}$,
F.~Menanteau$^{16,17}$,
R.~Miquel$^{45,18}$,
R.~L.~C.~Ogando$^{7,23}$,
A.~Palmese$^{8,28}$,
F.~Paz-Chinch\'{o}n$^{46,17}$,
A.~A.~Plazas$^{44}$,
A.~K.~Romer$^{47}$,
C.~S{\'a}nchez$^{48}$,
E.~Sanchez$^{24}$,
V.~Scarpine$^{8}$,
M.~Schubnell$^{2}$,
S.~Serrano$^{29,30}$,
M.~Smith$^{49}$,
M.~Soares-Santos$^{3}$,
E.~Suchyta$^{50}$,
G.~Tarle$^{3}$,
C.~To$^{31,12,13}$,
and R.D.~Wilkinson$^{47}$ 
\vspace{0.2cm} (DES Collaboration) \\
}

\ead{ben.henghes.13@ucl.ac.uk}
\vspace{10pt}
\begin{indented}
\item[] December 2020
\end{indented}

\begin{abstract}
In this paper we investigate how implementing machine learning could improve the efficiency of the search for Trans-Neptunian Objects (TNOs) within Dark Energy Survey (DES) data when used alongside orbit fitting. The discovery of multiple TNOs that appear to show a similarity in their orbital parameters has led to the suggestion that one or more undetected planets, an as yet undiscovered \enquote{Planet 9}, may be present in the outer Solar System. DES is well placed to detect such a planet and has already been used to discover many other TNOs. Here, we perform tests on eight different supervised machine learning algorithms, using a dataset consisting of simulated TNOs buried within real DES noise data. We found that the best performing classifier was the Random Forest which, when optimised, performed well at detecting the rare objects. We achieve an area under the receiver operating characteristic (ROC) curve, $(AUC) = 0.996 \pm 0.001$. After optimizing the decision threshold of the Random Forest, we achieve a recall of 0.96 while maintaining a precision of 0.80. Finally, by using the optimized classifier to pre-select objects, we are able to run the orbit-fitting stage of our detection pipeline five times faster.

\end{abstract}

%
\vspace{2pc}
\ioptwocol

\section{Introduction}

The idea that additional planets may be present in the outer solar system has existed in astronomers minds since the successful prediction and subsequent discovery of Neptune in 1846 \citep{le1839variations, galle1846account}. Indeed, the discovery of the once major planet Pluto came as a direct result of a rush to find further planets \citep{pluto}. After finding many other minor bodies in the outer solar system, the possibility of there still being a large planet left to discover seemed unlikely. However, recent detections of more Trans-Neptunian Objects (TNOs) have led to a resurgence in hunting for the elusive \enquote{Planet 9}.  

This rekindled excitement was caused by an observed similarity of orbital parameters of certain TNOs, and was first noted by \citet{VP113} in their detection of $2012 VP_{113}$. Objects like $2012 VP_{113}$ have higher eccentricities, inclinations, and orbit further from the sun than the majority of TNOs, giving them the name \enquote{Extreme-TNOs} (ETNOs). ETNOs typically have a semi-major axis, $a > 150 AU$, and a perihelion distance, $q > 30 AU$, and it was shown that these objects displayed a grouping with similar arguments of perihelia, $\omega \approx 0^{\circ}$, that could be explained by a large distant planet. The initial theory was that this planet caused the similar orbital elements via the Kozai mechanism \citep{kozai} whereby the oscillating argument of perihelia of the objects, about a value of either $\omega = 0^{\circ}$ or $\omega = 180^{\circ}$, would cause an exchange between the eccentricity and inclination of the body. However, this seemed improbable due to the lack of observations of TNOs with  $\omega = 180^{\circ}$ \citep{2014defm}. Instead it was suggested by \citet{B&B1} that the planet would cause similarities in both the argument of perihelia, $\omega$, and the longitude of ascending node, $\Omega$, through secular effects \citep{batygin2017dynamical}, which could then also account for other features seen in the Kuiper Belt \citep{batygin2016generation}. 

Having another major planet in the outer Solar System would result in other observable effects. Both \citet{gomes2016inclination} and  \citet{bailey2016solar} suggested that the six-degree Solar obliquity could be explained as a natural result of the additional planet. And, as discussed by \citet{cassini}, a Planet 9 with a true anomaly of $v \approx 120^{\circ}$ would significantly reduce the observed Cassini residuals. There have also been further studies which have reexamined the likeliness of an additional planet, with \citet{caceres2018influence} having suggested that in fact smaller perihelion distances provide better confinements.

However, there is still no consensus on whether the resonant perturbation mechanism is sufficient to describe the observed clustering of TNOs, what the most likely parameters of the planet are, or if it is in fact likely for there to be a planet at all \citep{batygin2019planet}. There have been several alternative proposals for how such a clustering of TNOs could be explained, ranging from regular secular dynamics being sufficient \citep{beust2016orbital}, to the possibility of a primordial Black Hole \citep{scholtz2019if} which could have been captured instead of a free-floating planet. Furthermore, there are other difficulties of explaining such a Planet 9 as it's thought to be unlikely to have migrated into its current orbit, or to have been a captured free-floating planet \citep{parker2017planet}. 

Finally, it is also uncertain if the grouping of TNOs that the entire Planet 9 hypothesis was based on is actually due to observational bias \citep{bernardinelli2020testing}. Using the Outer Solar System Origins Survey (OSSOS) \citep{bannister2016outer}, \citet{shankman2017ossos} discovered eight ETNOs which they claim have orbital parameter distributions consistent with what they would expect to detect and not grouped by a ninth Planet. Whereas \citet{brown2017observational} argues that the observed ETNOs must be grouped by external perturbations \citep{brown2019orbital}. \citet{sheppard2018new} also suggest that an additional planet is still favoured, but concede that more studies would need to be done which fully take into account the selection functions of the various surveys used to observe the ETNOs. It's essential to enhance the current search and discover more ETNOs to place further constraints on the Planet 9 hypothesis. However, regardless of the existence of Planet 9, more TNOs need to be discovered to better understand the structure of the outer Solar System.

The Dark Energy Survey (DES), while constructed as a cosmological survey, is perfectly situated to discover faint objects in the outer Solar System with its repeat observations in a 5,000 square degree footprint, and its ability to identify very dim objects with a $10\sigma$ limiting magnitude of $23.2$ in the r-band \citep{neilsen2016limiting} using its powerful camera, DECam \citep{flaugher2015dark}. Because a Planet 9 with a mass in the range $5 < M < 10M_e$ would have an aphelion magnitude of between $21.2 < V_{mag} < 24$ \citep{batygin2019planet}, it should be detectable within the DES footprint. Indeed, DES and DECam have already been used to discover many TNOs, including two of the ETNOs first used to hypothesise the presence of Planet 9 \citep{VP113, DES_more, gerdes2016observation, gerdes2017discovery, becker2018discovery, khain2018dynamical, khain2020dynamical, lin2019evidence, bernardinelli2020trans}. 

Our current process of detecting ETNOs and other distant objects in DES is to first combine observations of objects across different images. This is done by linking pairs of objects moving across images where the observed motion is consistent with Earth's parallax motion \citep{khain2020dynamical}. These pairs can then be joined to give sets of three points in three different images taken on three separate nights. With these sets of three points, (which we call triplets), it's then possible to fit them to an orbit to determine if there can be a bounded orbit well defined by the 6 parameters: Semi-major axis, $a$; Eccentricity, $e$; Inclination, $i$, Longitude of ascending node $\Omega$; Argument of perihelia, $\omega$; and the Mean Anomaly, $M$ \citep{bernstein2000orbit}.

Currently this process of orbit fitting is the slowest stage in our detection pipeline, and as the vast majority of triplets formed were from linking pixel-level fluctuations and artifacts that remained after difference imaging \citep{kessler2015difference} (which we refer to as noise), a lot of time is spent identifying this noise. In this paper we suggest an alternative method by implementing machine learning (ML), which is separate from the work done by \citet{bernardinelli2020trans}, and \citet{holman2018dwarf} who aimed to reduce the number of erroneous triplets that were initially linked. The machine learning classifier acts as an extra preprocessing stage to filter through the sets of triplets, identify and eliminate the majority of the triplets that result from noise in the data, and hence speed up the orbit fitting stage. The first step of this process is to train the ML algorithms on simulations of ETNO triplets created using a survey simulator. Simulations are necessary as there simply aren't enough real observations of these distant objects to form a sufficient training set; however, it is possible to combine these simulations with real noise data to ensure the training data is representative.

Eight different supervised ML algorithms are trained and tested, each contained in the Scikit-Learn python package \citep{scikit-learn}, and we find that for this task of classifying rare events, the Random Forest classifier (RF) is the best performing algorithm. Once optimised and implemented in the detection pipeline, the RF allows for $80\%$ of the noise triplets to be removed before performing orbit fitting which, as a result, runs five times faster.

In the following Sec 2. we describe the process of creating the simulated datasets that are used, and then summarise how to extract useful features. In Sec 3. we describe the ML algorithms tested and how the final classifier is optimised before giving the results in Sec 4. We then implement the classifier into the full search pipeline in Sec 5., and finally conclude our work in Sec 6.

\section{Data Simulations and Feature Extraction}

To be able to implement a supervised ML system, the ML algorithms first had to be trained and tested on data where the classifications were already known. As our problem focuses on finding rare events, there was not sufficient real data to form a large enough, and effective, training set to determine the algorithms' usefulness. Thus, synthetic objects were created and used to make simulated detections of ETNOs and possible Planet 9s within DES.

The simulations of detected ETNOs, and possible Planet 9s, were made using a survey simulator \citep{hamilton2019studies} which took an array of orbital parameters of objects given in Table~\ref{tab:orbital parameters}, and by using these fake orbits, calculated whether or not each object could ever be visible in DES. This was done by calculating the limiting magnitude of each exposure within $7^{\circ}$ of the position of the fake object at the beginning of DES operations (as even the fastest moving TNO couldn't have moved that far since the start of DES operations), and then project the position of each object into these nearby exposures to determine whether the object fell on a CCD during that exposure \citep{hamilton2019simulation}.

For the objects which could be detected, the simulator gave their positions in each DES image which could then be linked across multiple images. As the simulated objects were so distant, their motion across images was dominated by Earth's parallax motion, so pairs of objects could be found by linking the objects with motion consistent with the parallax motion. Pairs with common points were then combined to form triplets, sets of three points linked across three different images, as three points was the minimum requirement to perform an orbit fit to determine whether the observed points corresponded to an object or arose from noise in the images (see Sec 5. for a more complete description of the detection pipeline).

The majority of the dataset used was made up of real data which contained $\approx 250000$ triplets that had previously been linked but shown to result from noise after using the original method of performing a full orbit fit on every triplet detected. Although these real triplets could contain some small number of detections of objects which were miss-classified, the vast majority were confidently due to noise, and the machine learning algorithms used shouldn't have been noticeably impacted. The real data acted as the sea of negatives, in which we searched for the much rarer positive triplets, of which around $10000$ were made from the simulated objects. However, even with a far more noise triplets compared to the triplets from simulated objects, this imbalance in the dataset was still less than would be observed in real data where over 99.9\% of triplets result from noise. 

\begin{table}
\centering
\caption{Range of the 4 orbital parameters which were required by the Survey Simulator to create fake ETNOs and Planet 9s. In addition to these parameters the three further orbital parameters required to fully describe an orbit - Mean anomaly (M), argument of perihelion ($\omega$), and longitude of ascending node ($\Omega$), were also taken to have a uniform distribution. With these parameters the Simulator generated fake observations which could then be linked to generate the fake triplets used for the training data.}

	\label{tab:orbital parameters}
	\begin{tabular}{ccc}
		\hline
		 Parameter & Range  \\
 		\hline
		 Semi-major axis, $a$ & $ 150AU < a < 1000AU $  \\
		 Eccentricity, $e$ & $ 0.1 < e < 1 $ \\
		 Inclination, $i$ & $ 0^{\circ} < i < 90^{\circ} $ \\
		 Absolute Magnitude, $H$ & $ 1 < H < 10 $ \\
		\hline
	\end{tabular}
\end{table}

With the data prepared the next important step was feature extraction, whereby the features which were used by the ML algorithms were selected. In the case of having many raw parameters, one of the main aims of feature extraction is to lower the dimensionality of the data. There are several ways to reduce the dimensionality but perhaps the common way this can be done is by using the coefficients of principal component analysis (PCA) \citep{PCA} as features instead of features taken directly from the raw data. However, our specific problem had a very low dimensionality to begin with, and as such the process of feature extraction became more of a task to see what transformations could be made to the data to give the features which resulted in the best classifications. 

The raw data output by DES and the simulator contained the positions on the sky of each possible object in the image along with the time of observation. The most basic features which could be used were therefore the positions of the object in each image and the times of observation, giving a total of 9 features. However, instead of using the equatorial coordinates right ascension (RA) and declination (Dec), which were given as outputs in the raw data, it was found that by transforming the data into other coordinate systems the classifications could be greatly improved. As we were dealing with Solar System orbits it made more sense to use ecliptic coordinates to allow the ML algorithms to more easily infer whether or not the observations could result from a real orbit. 

Furthermore, instead of using the ecliptic positions of longitude and latitude of each of the three points with the times they were observed, as we were investigating moving objects, the positional data of individual points could be combined with their times to give the velocities between points in the triplet. And since we were interested in the overall motion, these velocities could be combined as in equations $(1) \& (2)$ to give changes in the velocities between each of the points.

\begin{equation}
dvlon = \frac{vlon_{12} - vlon_{23}}{vlon_{12} + vlon_{23}}
\label{eq:1}
\end{equation}

\begin{equation}
\centering
dvlat = \frac{vlat_{12} - vlat_{23}}{vlat_{12} + vlat_{23}}
\label{eq:2}
\end{equation}

\begin{figure}
    \centering
    \includegraphics[scale=0.15]{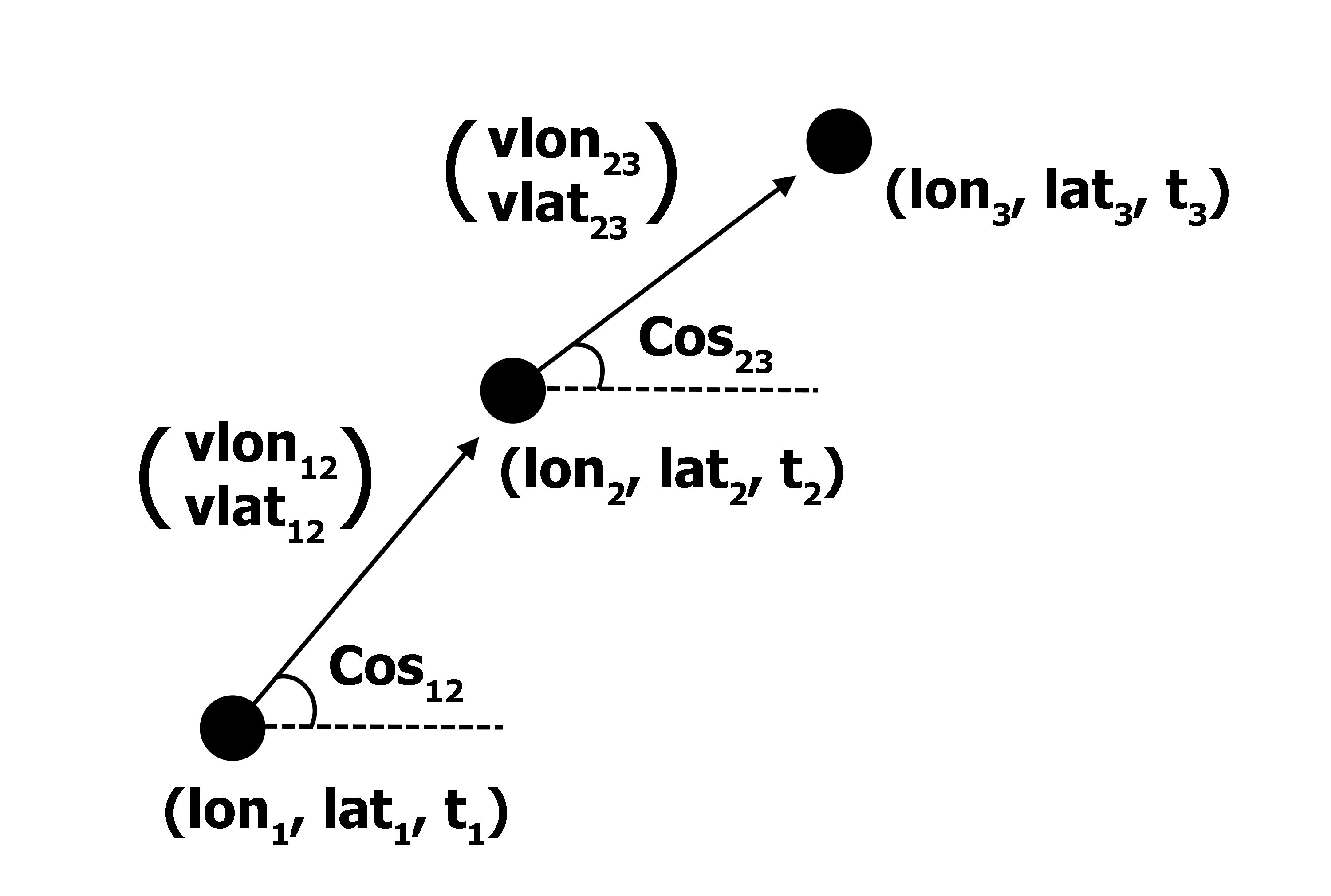}
    \caption{
    A triplet (displayed here to lie on a flat plane) was made by combining three points which had been linked across three different images taken on different nights. By transforming into ecliptic coordinates we were left with three sets of longitude and latitude as well as the times of the observations, resulting in 9 features. We then reduced the number of features by calculating the longitudinal and latitudinal velocities between each point, and further reduced these to simply the change in each velocity. By also using the two cosines between the two pairs of observations as features, we included all information needed about the trajectory of the object to be used by the machine learning algorithms to infer if the object could have a real orbit.}
    \label{fig:triplet}
\end{figure}

This reduced the initial 9 features from the coordinates down to just two, but to include all the information about the trajectory of the object, the cosines of the angles between points in the triplet were also included as features. This resulted in the final four features used by the ML algorithms: the change in longitudinal and latitudinal velocities (dvlon, and dvlat), and the cosine of the angles between points (cos$_{12}$, and cos$_{23}$), which is displayed in Figure \ref{fig:triplet} of a triplet.

\section{Machine Learning Methodology}

Having extracted the useful features, the next stage was to test various ML algorithms to determine which algorithm would give the best classification results. Here we perform tests on eight different supervised ML algorithms with the aim of implementing a new ML stage to increase the efficency of our TNO detection pipeline. Each of the classifiers, described below in section 3.1, were tested using the method described in section 3.2, with a full description of the optimisation performed given in section 3.3, and the metrics shown in section 3.4. Finally, the classification results are discussed in section 4.

\subsection{Description of classifiers}

\subsubsection{Naive Bayes}

Naive Bayes is a supervised algorithm that applies Bayes Theorem with the \enquote{naive} assumption that each pair of features is independent \citep{NB}. For the class variable~$y$, (in this case the label specifying whether the triplet results from an object or noise), and a dependent feature vector~$x_i$ (which here was the vector of the four features dvlon, dvlat, cos$_{12}$, and cos$_{23}$ described in Sec 2.), Bayes Theorem can be applied, where we use Maximum A Posteriori estimation to obtain an estimate for $P(x_i | y)$ and $P(y)$, where $P(y)$ is simply the relative frequency of class y in the training set.  
We tested a Gaussian Naive Bayes where the likelihood is taken to be Gaussian and the parameters $\mu_y$ and $\sigma_y$ are estimated using maximum likelihood estimation, and using these probabilities given by equation 3, we can then use Bayes Theorem to obtain the final probability of obtaining the class~$y$ given the features~$X$ which gives the classification result.

\begin{equation}
P(x_i \mid y) = \frac{1}{\sqrt{2\pi\sigma^2_y}} \exp\left(-\frac{(x_i - \mu_y)^2}{2\sigma^2_y}\right)
\end{equation}

\begin{equation}
P(y \mid X) = P(x_1 \mid y) \times P(x_2 \mid y) \times \dots \times P(x_i \mid y) \times P(y)
\end{equation}

\subsubsection{Logistic Regression}

Logistic Regression (LR) is a linear model used to make classifications, where the probabilities describing the outcome are modeled using the logistic function \citep{ESL}. As a linear model the target class variable~$y$ is assumed to be a linear combination of the features~$x_i$ with coefficients $\omega_i$, as given in equation 5, and the model fitting is analogous to least squares regression. 

\begin{equation}
y(w, x) = w_0 + w_1 x_1 + ... + w_i x_i
\end{equation}

\subsubsection{Multi-layer Perceptron}

A Multi-layer Perceptron (MLP) is an example of a deep neural network, consisting of at least three layers of nodes with the input, output and a minimum of one hidden layer. Similar to logistic regression, MLP learns a function to map the set of input features to the target vector; However, it differs from LR with the non-linear hidden layers which allow MLP to approximate any continuous function \citep{MLP}. In the input layer, each node represents a single feature. The following hidden layers then act to transform the previous layer using nodes which represent a different weighted linear summation of the input followed by a non-linear activation function. Finally the output layer takes the values from the previous hidden layer and transforms them into the output values. The weights of the linear summations are adjusted in training using backpropagation, where the gradient of the error function is calculated from the final layer backwards. The calculation of the gradient at each layer is reused in the computation of the gradient for the previous layer, and this backwards flow of information allows for more efficient computation of the gradient at each layer.

\subsubsection{k-Nearest Neighbours}

k-Nearest Neighbours (kNN) is an instance-based ML algorithm, whereby it doesn't attempt to make a general model used to classify data, but instead stores the training data which is used to classify new points. The classification is made by using a predefined number of points, k,  in the training sample which are closest to the new data point, and the classifier then predicts the class of the new point based on these neighbours \citep{knn}. Increasing the value of k will typically reduce the effects of the scatter of values, however, it will also make the classification boundary less distinct and can result in overfitting.

\subsubsection{Decision Trees}

Decision trees (DT) \citep{DT2} are non-parametric classifiers which work by using the data features to learn simple decision rules. These rules are basic if-then-else statements that are used to split the data into branches \citep{DT3}, and the tree is trained by recursively selecting the best feature split according to a pre-selected metric \citep{DT1}.  

While Decision trees can give a high accuracy, they aren't good at generalising the data, and are typically complex and overfitted to the training data. This could be improved by `pruning' the tree to make it a simpler model which could apply to more data, or by using an ensemble method which combines multiple decision trees to reduce overfitting in one of two ways: boosting or bagging.  

\subsubsection{Boosted Decision Trees}

Boosted Decision Trees (BDT) are the first ensemble method we considered, and multiple types of boosted classifiers were tested. The first was AdaBoost \citep{Adaboost} where the process of boosting is applied by repeatedly fitting the same dataset, but each time increasing the weights of incorrectly classified objects. This should result in a classifier that focuses more on the rarer cases. 

The second method of boosting used was Gradient Tree Boosting where the boosting is generalised by using an arbitrary differentiable loss function which is then optimised \citep{BDT1}. The loss function can be either binomial deviance or exponential, with the exponential loss function recovering the AdaBoost method; however, we found the deviance loss function to result in a better performing classifier for this problem. 

\subsubsection{Random Forests}

\begin{figure}
    \centering
    \begin{adjustwidth}{-0.7cm}{}
	\includegraphics[scale=0.31]{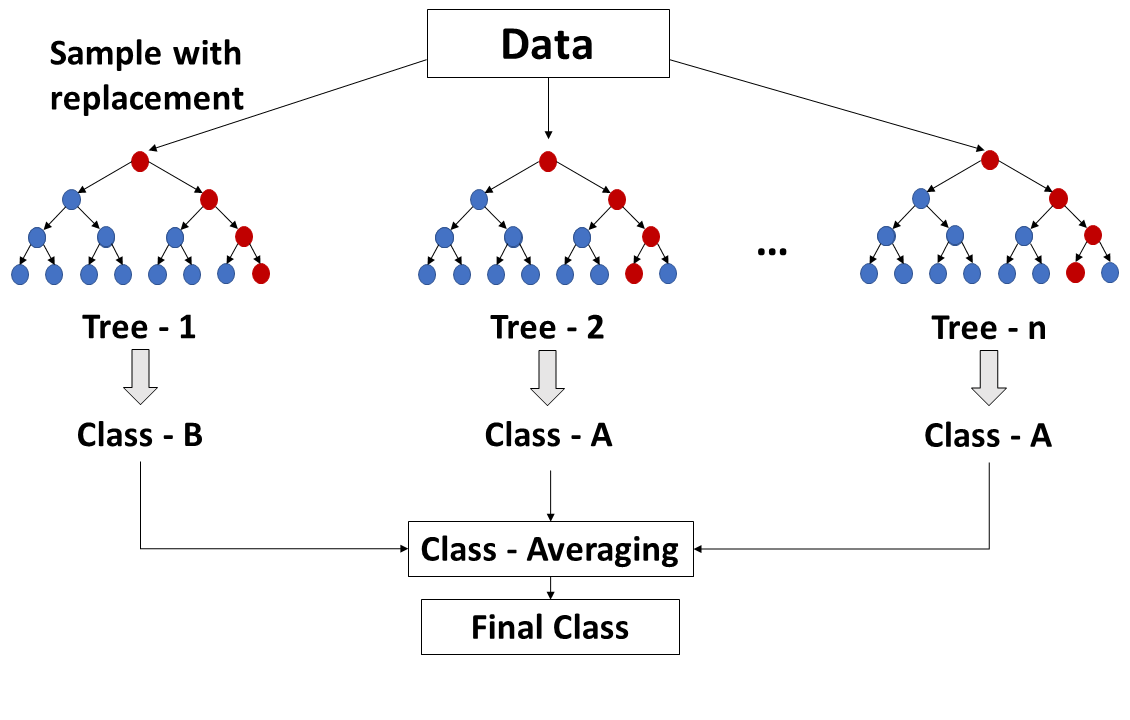}
	\end{adjustwidth}
    \caption{Flow chart showing the construction of a Random Forest. The data is sampled with replacement resulting in many decision trees being trained on random subsets of the data. A second element of randomness is then added where the feature splits performed in each tree are no longer the splits resulting in the highest information gain, but rather the best splits are taken from a random subset of the features. The class outputs of all the individual trees in the forest are averaged to give the final prediction, which has a much reduced variance and isn't as prone to overfitting as a single decision tree.}
    \label{fig:rand_forest}
\end{figure}

Random Forests (RF), such as the example in Figure \ref{fig:rand_forest} are another ensemble method that take many Decision Trees and averages their predictions \citep{RF1}. The forest is made by taking the entire dataset and sampling with replacement, giving random subsets of the data that are then used to construct many Decision Trees \citep{Bagging}. This creates an element of randomness where the dataset used to train each decision tree will be independent from every other tree making up the forest. Another element of randomness implemented by the RF is that the feature splits used aren't chosen by selecting the best possible split, and instead a random subset of the features is taken and the best split of these random features is used. As a result of this randomness, the bias (systematic error) of the RF usually increases compared with the bias of a single tree, however, after averaging all the trees, the variance decreases and more than compensates for this increase in bias. The result is a model which not only performs better but is also far less prone to overfitting. 

\subsubsection{Extremely Randomized Trees}

The Extremely Randomized Trees (ERT) \citep{ERT}, or \enquote{Extra Trees}, classifier was the final ensemble method tested. It is similar to the random forest with an extra step of randomness. When the feature splits are made, instead of using the most discriminative thresholds of the features, thresholds are picked at random for each feature and the best of these random thresholds is then used as the splitting decision rule. This usually acts to reduce the variance even more, at the expense of slightly greater bias.

\subsection{Methodology}

To test each of these supervised ML algorithms, the data first had to be split into training,  testing, and validation sets. We also performed cross-validation \citep{crossval} across the train/test set to select the best combination of hyperparameters, the parameters which are used to build the ML algorithms (see Sec 3.4 for a full description of the Optimisation), and to be able to identify and minimise the impact of any possible overfitting. We used a random split taking 70\% of the data to be used to train the algorithms, giving a healthy training set size with over 200000 triplets, and leaving the remaining 30\% for tests and validation. However, due to the nature of the problem of looking for rare events, the datasets were incredibly imbalanced which could cause problems when trying to perform the classifications. Rather than forcibly making balanced datasets, which then wouldn't be representative of the true data, certain algorithms (LR, DT, RF, ERT) could take into account the class imbalance by applying a weighting to the data during the training. 

Another common step taken before training any algorithm is to scale the data, making all the features have a range $0 < |feature| < 1$. This can be useful if the ML algorithm uses gradient descent, as convergence will occur much faster on normalised data \citep{sgd}. It can also be necessary if the data features have different units and varying ranges of values, as some models are sensitive to magnitude or use the euclidean distance between points such as kNN. However, scaling could also remove useful information if the difference in ranges of the features is important or resulting from some physical effect. Furthermore, scaling may not even be possible if the full range of data isn't known. In these tests, although scaling was also applied, it was found to make very little difference to the performance of the majority of the classifiers and was ignored when implementing the final RF classifier. 

\subsection{Hyperparameter Optimisation}

The final step of creating the ML algorithms to be tested was to state the hyperparameters. Hyperparameters are the parameters of ML algorithms that get set prior to the learning process, and are used to create the algorithms, allowing all other parameters of the model to be learned from the training data. As an example, in random forests these hyperparameters include the number of estimators, which is the number of trees that make up the forest.

We tested three different methods of optimisation each of which required a grid of hyperparameters to be defined. This grid defines the hyperparameters and the range of values to be tested. The final thing needed to perform optimsation is the metric that is being optimised for. As described in the following section, the recall was the most informative metric for this problem of searching for rare objects and therefore we optimised the algorithms to give the best possible recall scores. 

The first method, `Brute Force Optimisation', is a method where all possible combinations of the hyperparameters in the grid are tested. While this is a computationally expensive task, it is the most complete way to optimise a ML algorithm. The next method used was `Random Optimisation' in which a predefined number of iterations were performed with a random selection of the hyperparameters from the grid. This method is far faster than testing every combination in the hyperparameter grid, and although it will usually give a slightly worse result, it's only a small difference when compared to the amount of time saved. Similarly the final method, `Bayesian Optimisation', iterates until it converges on the best set of hyperparameters \citep{bayesopt}. This method is a good middle ground, slightly more thorough than a completely random search, but nowhere near as exhaustive as the brute force method.

For the initial tests of all of the ML algorithms a simple brute force search was completed on the grids defined in Table~\ref{tab:all_classifiers_grid}. These values of hyperparameters were chosen to provide a wide enough range and ensure sufficient variation for each algorithm by changing the values of the hyperparameters which had the greatest effect.  Although some hyperparameters could have continued to been increased past the chosen upper limits, such as the number of estimators used in the ensembles of decision trees, we used a maximum which would provide a good estimate of performance without taking days to compute. Similarly, rather than testing every hyperparameter, to save time we only selected the ones with the greatest impact on the algorithm and the remaining hyperparameters not listed in Table~\ref{tab:all_classifiers_grid} were kept at their default Scikit-Learn values. For a complete analysis benckmarking would be required to fully understand the trade off between the training and inference times and the accuracies obtained, however, this was beyond the scope of this paper.

After the results of these tests were obtained (which are shown in Table~\ref{tab:test_results} in Sec 4), and the Random Forest was selected as the best performing classifier, a more complete optimisation was performed. All three optimisation methods were tested on the larger grid given in Table~\ref{tab:optimisation_grid} and the results are given in Table~\ref{tab:optimisation_results}. While all three optimisation techniques were successful in improving the performance, the Brute force method did give the largest improvements. However, the differences were minimal, and the factor of 10 difference in the time taken compared to the other methods makes using one of the alternative methods more appealing. Furthermore, while changing the hyperparameters does fine tune the algorithm and improve classification results, the effect is far less than changing the data itself and to improve the results any further one would need to add features in the data processing stages.

\begin{table}
\centering
\caption{Grids of hyperparameters that were searched when constructing each classifier for the initial tests to be able to compare each of the machine learning algorithms. Hyperparameters not mentioned in the table were kept at the default Scikit-Learn values. The hyperparameters that were selected to be used for each classifier are shown in bold, with the exception of NB in which the only hyperparameter which can be set are the prior probabilities which were left to be automatically adjusted according to the data.}

	\label{tab:all_classifiers_grid}
    \begin{adjustbox}{width=\columnwidth,center}
	\begin{tabular}{ccc}
		\hline
		 Classifier & Hyperparameter & Array of Values  \\
 		\hline
    	   LR & ``dual" & [False, \textbf{True}]\\
              & ``tol"& [\textbf{1e-7}, 1e-6, 1e-5, 1e-4] \\
              & ``C" & [1.0, 2.0, 3.0, \textbf{4.0}, 5.0]\\
		\hline
		kNN & ``no. neighbors" & [\textbf{1}, 5, 10, 50, 100]\\
             & ``weights" & [\textbf{``uniform"}, ``distance"]\\
             & ``leaf size" & [\textbf{1}, 5, 10, 50]\\
		\hline
		DT & ``min. samples split" & [\textbf{2}, 5, 10, 50]\\
              & ``criterion" & [``gini", \textbf{``entropy"}]\\
              & ``splitter" & [\textbf{``best"}, ``random"]\\
        \hline
		BDT & ``loss" & [``exponential", \textbf{``deviance"}]\\
              & ``no. estimators" & [50, 100, 150, \textbf{200}]\\
        \hline
		RF & ``no. estimators" & [10, 50, 100, \textbf{200}]\\
              & ``max. features" & [\textbf{``auto"}, 0.1, 0.4]\\
              & ``min. samples leaf" & [\textbf{1}, 5, 10, 20]\\
        \hline
		ERT & ``no. estimators" & [10, 50, 100, \textbf{200}]\\
              & ``max. features" & [\textbf{``auto"}, 0.1, 0.4]\\
              & ``min. samples leaf" & [\textbf{1}, 5, 10, 20]\\
        \hline
		MLP & ``hidden layer sizes" & [\textbf{1}, 10, 50, 100]\\
              & ``tol" & (\textbf{1e-3}, 1e-4, 1e-5)\\
        \hline
	\end{tabular}
	\end{adjustbox}
\end{table}

\begin{table}
\centering
\caption{Grid of hyperparameters used by the three different techniques in the optimisation process for the Random Forest. For the Random and Bayesian optimisations only the upper and lower values were used to obtain a random value between the two limits, whereas for Brute force optimisation the specific values within the range also had to be stated. Additional hyperparameters not listed in the table were kept at the default Scikit-Learn values. The final hyperparameter values which gave the highest recall score are given in bold.}
	\label{tab:optimisation_grid}
	\begin{tabular}{cc}
		\hline
		 Hyperparameter & Array of Values  \\
 		\hline
		 ``no. estimators" & [1, 10, 50, 100, \textbf{200}]\\
		 ``criterion" & ['gini', \textbf{'entropy'}]\\
		 ``max. features" & [\textbf{0.1}, 0.4, 0.9]\\
		 ``min. samples split" & [\textbf{2}, 5, 10, 20]\\
		 ``min. samples leaf" & [\textbf{1}, 5, 10, 20]\\
		 ``min. weight fraction leaf" & [\textbf{0}, 0.4]\\
		 ``bootstrap" & [True, \textbf{False}]\\
		\hline
	\end{tabular}
\end{table}

\begin{table}
\centering
	\caption{Results from using the random forest classifier when optimised using the three different methods as compared to the default classifier given by Scikit-Learn.}
	\label{tab:optimisation_results}
	\begin{adjustwidth}{-0.7cm}{}
	\begin{tabular}{|p{2cm}|p{1.3cm}|p{1.3cm}|p{1.3cm}|p{1.3cm}|}
	\hline
	 Optimisation technique & None & Brute Force & Random & Bayesian\\
	\hline
	Time Taken (s) & 0.00 & 43437.40 & 4654.01 & 7272.07 \\
	\hline
	Accuracy & 0.9891 $\pm$0.0004 & 0.9912 $\pm$0.0004 & 0.9907 $\pm$0.0003 & 0.9903 $\pm$0.0004  \\
	\hline
	Recall & 0.8588 $\pm$0.0061 & 0.9000 $\pm$0.0062 & 0.8976 $\pm$0.0057 & 0.8965 $\pm$0.0060  \\
	\hline
	Precision & 0.9129 $\pm$0.0096 & 0.9265 $\pm$0.0063 & 0.9139 $\pm$0.0083 & 0.9085 $\pm$0.0085\\
	\hline
	F1 score & 0.8847 $\pm$0.0023 & 0.9122 $\pm$0.0042 & 0.9055 $\pm$0.0014 & 0.9034 $\pm$0.0037\\
	\hline
	AUC & 0.9877 $\pm$0.0026 & 0.9963 $\pm$0.0011 & 0.9947 $\pm$0.0009 & 0.9930 $\pm$0.0011 \\
	\hline
	\end{tabular}
	\end{adjustwidth}

\end{table}

\subsection{Metrics}

Once the classifiers had been trained and tested their performance had to be determined. There are various metrics that can be used for analysing ML algorithms, the simplest of which is the classification accuracy. Although using the accuracy gave a quick way to determine how well a classifier performed, it wasn't particularly useful in the information it provided. As accuracy is simply the number of true predictions / total predictions (as defined in eq.\ref{eq:acc}), and in this case the majority of the data should be easily identified as a true negative, the null accuracy (predicting everything to be a negative result) was very high at 95\%. This means that quoting an accuracy which sounds incredibly good can in fact still be a very poor classifier, as seen in some of our tests. Instead of using the accuracy, far more useful metrics can be obtained from the confusion matrix, a matrix of the true values against the predicted values \citep{metrics}. 

The confusion matrix, such as the binary example in Table~\ref{tab:confusion_matrix}, allows for the other important metrics to be calculated from the number of true positives (TP), true negatives (TN), false positives (FP), and false negatives (FN). The most useful metric in this case was the recall (or completeness / sensitivity).

\begin{equation}
Accuracy = \frac{TP + TN}{TP + TN + FP + FN} 
\label{eq:acc}
\end{equation}

\begin{table}
    \centering
	\caption{Confusion matrix for a binary classification problem with two possible classes, positive (P) or negative (N).}
	\label{tab:confusion_matrix}
	\begin{tabular}{lccc}
		\hline
		 & & \multicolumn{2}{c}{Predicted Class} \\
		 & & Positive & Negative\\
		\hline
		True Class & Positive & TP & FN \\ 
		& Negative & FP & TN\\
		\hline
	\end{tabular}
\end{table}

\begin{equation}
Recall = \frac{TP}{TP + FN}    
\end{equation}

The recall gives the best measure of how many possible observations would be missed. For this problem we didn't want to have any FN which could have actually occurred due to a real object, and focused on optimising for the recall. However, improving the recall score came at the cost of decreasing the precision (or purity). And although we allowed for more FP, the precision had to also be kept as high as possible to not have too many FP which would make the machine learning method inefficient. 

\begin{equation} 
Precision = \frac{TP} {TP + FP}
\end{equation}

A combination of these two metrics, the F1 score, was used to show the balance between the recall and precision. The F1 score is the harmonic average of the two metrics, and as such also has its best value at 1 and worst at 0.

\begin{equation} 
F1 = 2\times \frac{Precision \times Recall} {Precision + Recall} 
\end{equation}

\begin{figure*}
    \centering
	\includegraphics[scale=0.75]{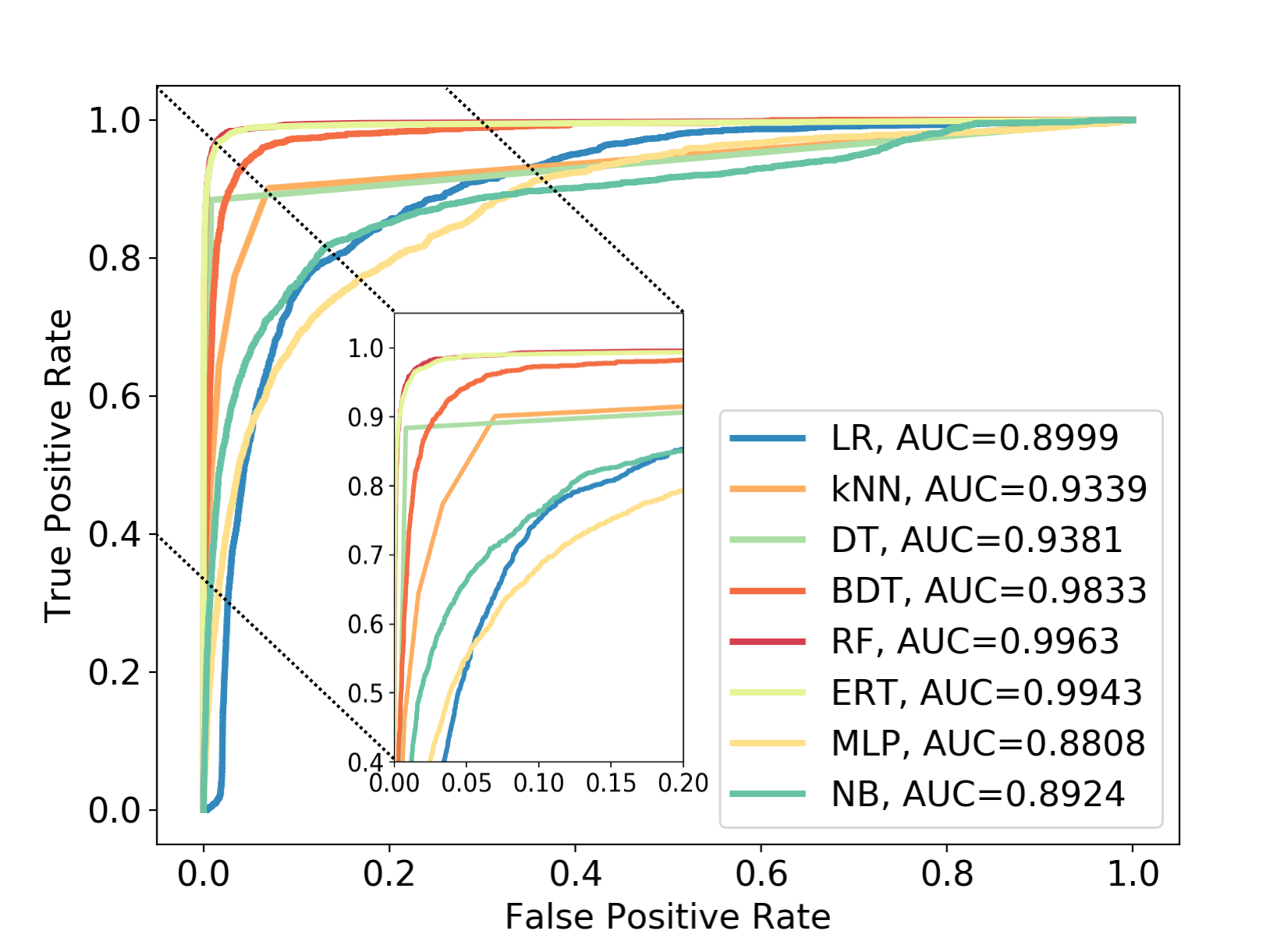}
    \caption{ROC curves for each machine learning classifier tested overlaid to be able to compare their effectiveness. The algorithms compared are: Logistic Regression (LR), K-Nearest Neighbours (kNN), Decision Trees (DT), Boosted Decision Trees (BDT), Random Forests (RF), Extra-Randomised Trees (ERT), Multi-layer Perceptron (MLP), and Naive Bayes (NB). The tree-based classifiers outperformed all others, with the Random Forest (RF) and Extra-Randomised Trees (ERT) being the best. The decision tree classifier (DT) produced a three-point curve as it outputs only the label rather than a predicted probability, giving only a single point of interest to plot.}
    \label{fig:ROC_tests}
\end{figure*}

Another very useful metric is the area under the curve (AUC) of the receiver operating characteristic curve (ROC curve) \citep{roc}. The ROC curve was plotted with the True positive rate (TPR) against the false positive rate (FPR), resulting in a curve where the ideal result with AUC = 1 would be a straight line up and across. A ROC curve with each of the tested algorithms is shown in Figure~\ref{fig:ROC_tests}.

\begin{equation} 
TPR = \frac{TP} {TP + FN}
\end{equation}

\begin{equation} 
FPR = \frac{FP} {FP + TN}
\end{equation}

\section{Classification Results}

\begin{figure*}
    \centering
    \begin{adjustwidth}{-1cm}{}
	\includegraphics[scale=0.6]{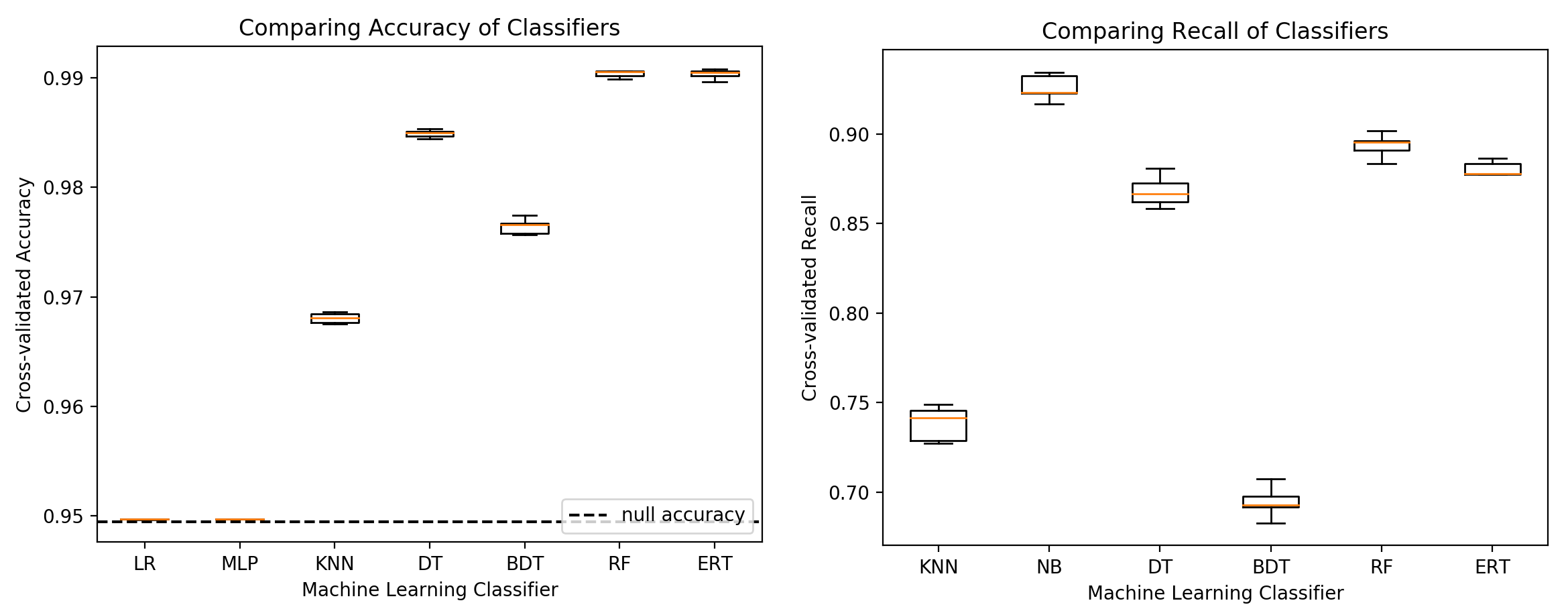}
	\end{adjustwidth}
    \caption{Box and Whisker plots comparing the accuracy and recall of the various different machine learning classifiers tested. The NB classifier is excluded from the accuracy graph as its value was so low ($\approx0.5$), and similarly both LR and MLP are removed from the recall plot as their recalls were essentially 0 having recovered the null result of predicting all triplets to result from noise.
    The interquartile range (making up the box) was obtained by performing cross-validation in the tests of every algorithm, which also provided a standard deviation to be used as the uncertainty. The range of the results was shown by the extended 'whiskers', and the median is shown in red.}
    \label{fig:BW_plot}
\end{figure*}

The full results of the tests are given in  Table~\ref{tab:test_results} and all of the quoted results were obtained using 5-fold cross-validation to obtain a value which was unaffected by overfitting, and allowed for the standard deviation to be calculated. The relative usefulness is displayed in Figure~\ref{fig:ROC_tests}, a single plot overlaying the ROC curves for each of the classifiers, as well as using box and whisker plots in Figure~\ref{fig:BW_plot}.

\begin{table*}
\centering
	\caption{Results from testing the eight different machine learning algorithms described in section 4.1, the metrics were obtained using 5-fold cross-validation which also allowed for the standard deviation to be calculated and is given as the error.}
	\label{tab:test_results}
	\begin{tabular}{|p{1.5cm}|p{1.5cm}|p{1.5cm}|p{1.5cm}|p{1.5cm}|p{1.5cm}|p{1.5cm}|p{1.5cm}|p{1.5cm}|}
	\hline
	Classifier & Logistic Regression (LR) & K-Nearest Neighbours (kNN) & Naive Bayes (NB) & Decision Tree (DT) & Boosted Decision Tree (BDT) & Random Forest (RF) & Extra Randomised Trees (ERT) & Multi-layer Perceptron (MLP) \\
	\hline
	Accuracy & 0.946 $\pm$0.002 & 0.968  $\pm$0.001 & 0.503  $\pm$0.066 & 0.985 $\pm$0.001 & 0.976 $\pm$0.001 & 0.990 $\pm$0.001 & 0.990 $\pm$0.001 & 0.949 $\pm$0.001 \\
	\hline
	Recall & 0.001 $\pm$0.001 & 0.738 $\pm$0.009 & 0.926 $\pm$0.007 & 0.866 $\pm$0.008 & 0.694 $\pm$0.007 & 0.892 $\pm$0.006 & 0.880 $\pm$0.005 & 0.000 $\pm$0.000 \\
	\hline
	Precision & 0.018 $\pm$0.009 & 0.664 $\pm$0.005 & 0.088 $\pm$0.011 & 0.838 $\pm$0.005 & 0.811 $\pm$0.012 & 0.914 $\pm$0.007 & 0.924 $\pm$0.006 & 0.000 $\pm$0.000 \\
	\hline
	F1 score & 0.002 $\pm$0.001 & 0.690 $\pm$0.004 & 0.160 $\pm$0.019 & 0.852 $\pm$0.005 & 0.747 $\pm$0.007 & 0.902 $\pm$0.002 & 0.902 $\pm$0.004 & 0.000 $\pm$0.000 \\
	\hline
	AUC & 0.911 $\pm$0.006 & 0.859 $\pm$0.004 & 0.899 $\pm$0.007 & 0.929 $\pm$0.003 & 0.984 $\pm$0.001 & 0.996 $\pm$0.001 & 0.995 $\pm$0.001 & 0.866 $\pm$0.006 \\
	\hline
	\end{tabular}
\end{table*}

From these it's clear that some of the ML algorithms have completely failed to identify the triplets resulting from the fake objects. MLP found no TPs, and achieved the same as the null result of classifying everything as negative (triplets resulting from noise). LR performed similarly, classifying almost all triplets as negatives, and by falsely classifying some noise as positives (triplets resulting from simulated objects), it had an accuracy lower than that of the null accuracy. While it's possible that spending more time optimising these algorithms could have resulted in improving them to no longer give the null classification, they also would have continued to classify too many FPs and had a precision too low to improve the efficiency of the search pipeline. Furthermore, compared with the other algorithms which were able to provide much better results with only the quick optimisation which was carried out on all the algorithms, they were not worth considering for this task. 

The remaining classifiers all did much better, not having recall and precision scores close to 0, however, the tree-based classifiers seemed to be the best performing algorithms. Although kNN was somewhat successful, it had both a lower accuracy and precision/recall than most of the tree based methods, and didn't seem to be an optimal classifier for this problem. NB performed better than all other classifiers for the recall, but while this was the most important metric, it was only able to achieve this high a value by classifying almost half the data as positives, and as such it had a very low precision and by far the worst accuracy. The accuracy and precision being so low meant that it wasn't a useful classifier on its own, as it wouldn't be at all efficient when searching for objects; however, it could have been used if combined with other algorithms in a voting system, but this possibility is left for future work where we would consider more complex algorithms.  

The tree-based classifiers were strong performers, but with some crucial differences between them. The basic DT classifier, although did well classifying the training set, was slightly overfitting despite the cross-validation and had lower metric scores which meant that it wouldn't be useful when applied to new data. The ensembles methods were much better at addressing this overfitting, but the BDT was consistently worse than the randomised methods due it not being able to handle the huge imbalance between classes. As a result, in all metrics, the boosted trees did worse than both the DT and forest classifiers, making it more similar to kNN in performance and also not useful for this problem. There was less to distinguish between the RF and ERT classifiers which had very similar metrics and performed very well at classifying the rare events; however, the RF was the faster method taking almost half the time to train and complete the classifications. On top of this the RF had a higher recall, suggesting that the additional stage of randomness in ERT was unnecessary for this problem. 

\begin{figure*}
    \centering
    \begin{adjustwidth}{-0.1cm}{}
	\includegraphics[scale=0.2]{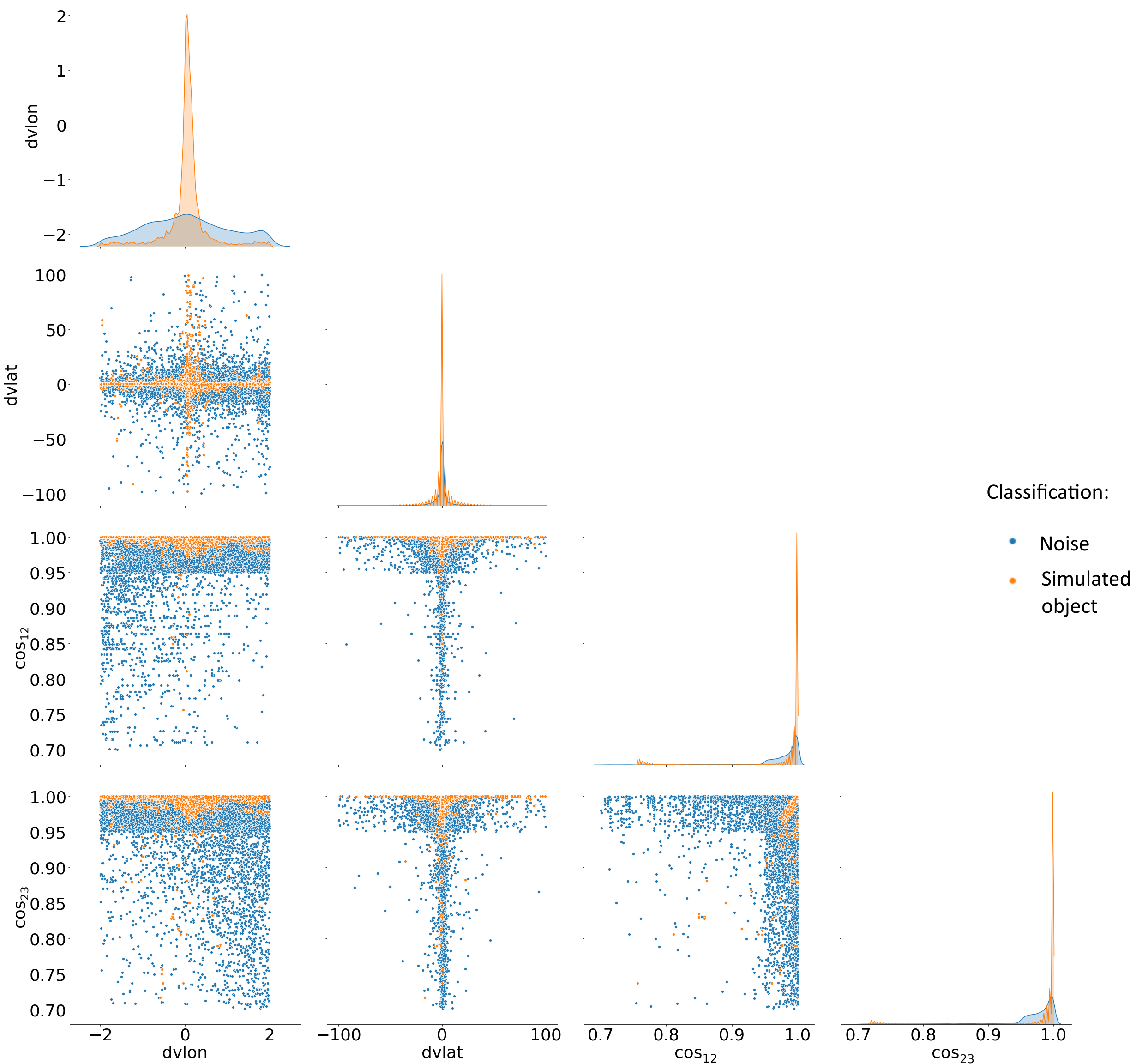}
	\end{adjustwidth}
    \caption{Pair plots showing the 4 features: dvlon, dvlat, cos$_{12}$, and cos$_{23}$, which were used by the Random Forest classifier plotted against each other with the label of their classification shown by the colour - blue for noise and orange for simulated objects. Machine learning was especially beneficial given the overlap of the two classes in each feature space meaning that there were no simple cuts able to separate the classes, and the Random Forest performed impressively, accurately classifying the majority of the triplets.}
    \label{fig:pair_plots}
\end{figure*}

Having selected the Random Forest as the most successful classifier, we then produced pair plots shown in figure \ref{fig:pair_plots} to examine the distributions of the features and suggest how the algorithm was able to produce its classifications. The majority of the simulated objects had quite sharp peaks due the fact that TNOs were more likely to have very small changes in longitudinal and latitudinal velocities and have cosines close to 1. Although one could have therefore used simple cuts to select the object closest to the peaks, by doing so far more triplets resulting from simulated objects would have been miss classified resulting in more possible detections getting missed. Instead, by implementing a machine learning algorithm like the Random Forest it was possible to achieve far better classifications, and the tree-based algorithms might have performed better than others due to their nature of using many decision rules to be able to `pick out' the majority of the simulated objects without also miss-classifying much of the noise. 

The final step taken to improve the performance of the Random Forest classifier was to change the decision threshold. In making classifications the RF calculated a predicted probability for each triplet to determine the probability of it resulting from noise or a real object. The decision threshold is the value at which the threshold is set so that for probabilities above this threshold the classification is taken to be positive (and the triplet results from a real object), and for probabilities below this threshold the classification is negative (and the triplet results from noise).

\begin{figure}
    \centering
    \begin{adjustwidth}{-1cm}{}
	\includegraphics[scale=0.5]{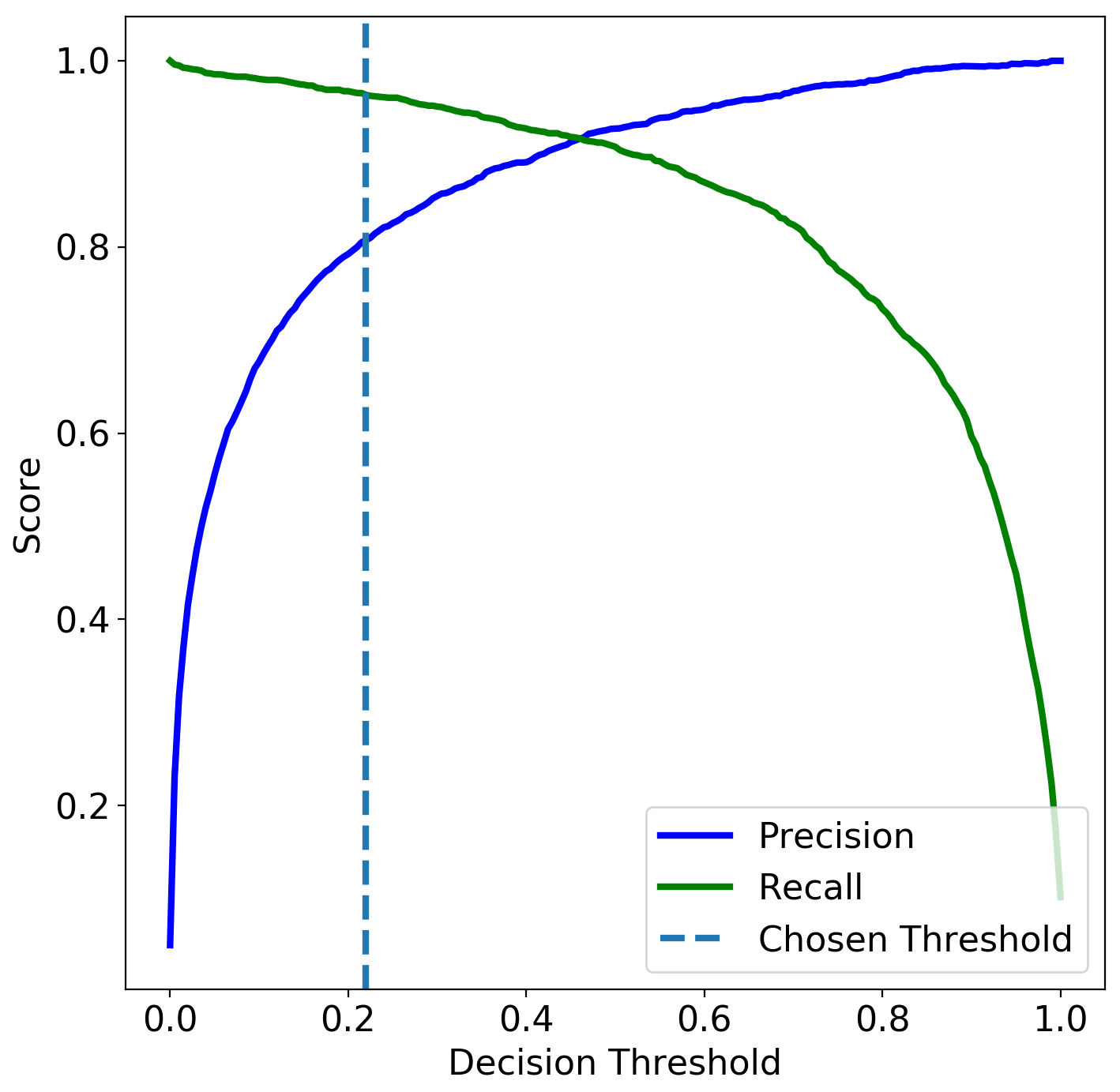}
	\end{adjustwidth}
    \caption{Graph showing the precision and recall scores of the Random Forest classifier as a function of the decision threshold. The default value of the threshold was set $= 0.5$, however, by changing to the chosen value $\approx 0.2$ we were able to achieve a better recall without reducing the precision (and hence efficiency) by too much.}
    \label{fig:decsion_threshold}
\end{figure}

The default threshold was set $= 0.5$, however, as can be seen in Figure~\ref{fig:decsion_threshold}, which shows how the recall and precision change with the decision threshold, we were able to obtain a better result for the recall by lowering this threshold. Although lowering the threshold to our chosen value $\approx 0.2$ resulted in a lower accuracy and precision, the recall improved sufficiently so that we were far less likely to miss a possible detection of a real object. Before changing the decision threshold the RF was missing 163 out of the 4600 (3.5\%) triplets from simulated objects that were in the test set. Having changed the threshold, this was lowered to only 73/4600 (1.5\%) of the triplets resulting from simulated objects being miss-classified as noise, and although this does mean missing these triplets, in the full pipeline multiple triplets from the same object are almost always required to actually result in a confirmed detection. As such, although some triplets were missed, enough triplets were correctly classified that the vast majority of real objects would still be recovered. 

\section{Detection Pipeline}

After the Random Forest had been found to be the best classifier, optimised, and had the decision threshold changed to further improve its performance, it was possible to implement the RF into the full search pipeline. 

Our pipeline to identify TNOs can be described in three main stages which we summarise here, and a complete description of the entire process was done by \citet{bernardinelli2020trans}. First, the observational data had to be linked to give the sets of observations that could be of the same object. For each point in the data a linkmap was used to produce an array of all possible points that could be linked to it, determined by whether or not the motion between points seemed to be consistent with Earth's parallax motion. For TNOs the motion needs to be consistent with Earth's parallax as they are such distant objects that their proper motion is much less apparent than the motion of the Earth. The output of the linkmap results in pairs of points that could possibly be the same object, and the next step was to take the linked pairs and form triplets, the sets of three points that could all be from the same object. This was done in the same way that the pairs were formed, checking to see if the motion from one set of pairs to the next was consistent. 

Once the triplet was formed it needed to be checked to see if it could have actually arisen from an object or if it was an artifact of noise in the data \citep{kessler2015difference}. This was where the ML classifier was implemented as an extra preprocessing step to quickly discard the majority of the triplets which result from noise in the data. After the majority of the noisy triplets had been removed (over 80\%), the remaining triplets were fitted to an orbit to see if they could be described by a real orbit of an object, or if they were still due to noise. This orbit fitting stage determined whether there could be a bounded orbit well described by the six orbital parameters: Semi-major axis, $a$; Eccentricity, $e$; Inclination, $i$, Longitude of ascending node $\Omega$; Argument of perihelia, $\omega$; and the Mean Anomaly, $M$ \citep{bernstein2000orbit}, and was the slowest stage of the pipeline. By removing most of the noise using the ML classifier rather than orbit fitting every triplet generated during linking, most of this computationally expensive step was saved, and the search was sped up significantly. 

The increase in efficiency by implementing ML was evident and the pipeline was run five times faster when using the classifier. This was achieved as, out of the dataset containing around a quarter of a million triplets, only around 10\% were kept to be fitted to an orbit, but even after keeping so few of the initial triplets, the classifier only miss-classified 73/4600 (1.5\%) triplets from simulated objects. Although this is still missing 1.5\% of the triplets arising from objects, real objects will be almost always be discovered by multiple triplets. As such having a recall $\approx 0.96$ which was achieved by the Random Forest is likely to recover the vast majority of the real objects. This would allow for the edited pipeline which includes the classifier to be run as a quick preliminary search and still be able to detect most of the objects before completing a full analysis.

\section{Summary}

The classification of rare events, such as this example of searching for ETNOs and a possible ninth Planet, has become an even more important venture in light of the vast datasets becoming available. In the wake of future surveys like the Vera C. Rubin Observatory (or Large Synoptic Survey Telescope (LSST)), which will produce 10 million transient events every night, being able to utilise ML methods will be vital to improve efficiencies and allow further analysis to be undertaken. 

In this work we've shown that implementing ML classifiers using the very user friendly package scikit-learn could be used as a preprocessing step, removing the vast majority of erroneous detections, which helped speed up our discovery pipeline. Having tested eight of the most used algorithms we discovered that the Random Forest classifier was the best performing overall, and had the best functionality of being less prone to overfitting and taking into account imbalanced datasets.

Our results showed that the optimised Random Forest used could perform incredibly well, and achieved an AUC = 0.996. Furthermore, by changing the decision boundary we maximised the recall, giving a recall $= 0.96$ to ensure that the vast majority of the triplets resulting from real objects could be recovered. We also maintained a high accuracy and precision at 0.99 and 0.80 respectively. This meant that our method was far more efficient, preventing the vast majority of the triplets resulting from noise from advancing to the orbit fitting stage, and greatly speeding up the pipeline. 

If used in parallel with the existing pipeline which fits all triplets to an orbit to ensure it's 100\% complete, implementing machine learning could allow for a useful preliminary search to identify objects more quickly and provide a cross check for the objects passing the orbit fitting.

The work presented here opens the door for analyses on searching for other populations of TNOs in DES data. This method of using machine learning to filter noise could be especially useful to help identify closer objects where the faster motion results in even more noise. It would be desirable to investigate whether the RF classifier would be as effective when applied to these different populations of objects, and implement a ML method at a similar stage in the detection pipeline.

Further investigation could also be done to implement new algorithms which have the potential to speed-up the pipeline even more, as well as using machine learning in other areas, such as changing the way that points can be linked through images, which will make it possible to further improve the current search. Improvements such as these will aid the discovery of far more of the TNO population which is crucial information for constraining Planet 9 and learning more about our Solar System.  

\section*{Acknowledgements}

BH was supported by the STFC UCL Centre for Doctoral Training in Data Intensive Science (grant number ST/P006736/1).

OL  acknowledges support from a European Research Council Advanced Grant TESTDE FP7/291329 and an STFC Consolidated Grants ST/M001334/1 and ST/R000476/1.

Funding for the DES Projects has been provided by the U.S. Department of Energy, the U.S. National Science Foundation, the Ministry of Science and Education of Spain, 
the Science and Technology Facilities Council of the United Kingdom, the Higher Education Funding Council for England, the National Center for Supercomputing 
Applications at the University of Illinois at Urbana-Champaign, the Kavli Institute of Cosmological Physics at the University of Chicago, 
the Center for Cosmology and Astro-Particle Physics at the Ohio State University,
the Mitchell Institute for Fundamental Physics and Astronomy at Texas A\&M University, Financiadora de Estudos e Projetos, 
Funda{\c c}{\~a}o Carlos Chagas Filho de Amparo {\`a} Pesquisa do Estado do Rio de Janeiro, Conselho Nacional de Desenvolvimento Cient{\'i}fico e Tecnol{\'o}gico and 
the Minist{\'e}rio da Ci{\^e}ncia, Tecnologia e Inova{\c c}{\~a}o, the Deutsche Forschungsgemeinschaft and the Collaborating Institutions in the Dark Energy Survey. 

The Collaborating Institutions are Argonne National Laboratory, the University of California at Santa Cruz, the University of Cambridge, Centro de Investigaciones Energ{\'e}ticas, 
Medioambientales y Tecnol{\'o}gicas-Madrid, the University of Chicago, University College London, the DES-Brazil Consortium, the University of Edinburgh, 
the Eidgen{\"o}ssische Technische Hochschule (ETH) Z{\"u}rich, 
Fermi National Accelerator Laboratory, the University of Illinois at Urbana-Champaign, the Institut de Ci{\`e}ncies de l'Espai (IEEC/CSIC), 
the Institut de F{\'i}sica d'Altes Energies, Lawrence Berkeley National Laboratory, the Ludwig-Maximilians Universit{\"a}t M{\"u}nchen and the associated Excellence Cluster Universe, 
the University of Michigan, NFS's NOIRLab, the University of Nottingham, The Ohio State University, the University of Pennsylvania, the University of Portsmouth, 
SLAC National Accelerator Laboratory, Stanford University, the University of Sussex, Texas A\&M University, and the OzDES Membership Consortium.

Based in part on observations at Cerro Tololo Inter-American Observatory at NSF’s NOIRLab (NOIRLab Prop. ID 2012B-0001; PI: J. Frieman), which is managed by the Association of Universities for Research in Astronomy (AURA) under a cooperative agreement with the National Science Foundation.

The DES data management system is supported by the National Science Foundation under Grant Numbers AST-1138766 and AST-1536171.
The DES participants from Spanish institutions are partially supported by MICINN under grants ESP2017-89838, PGC2018-094773, PGC2018-102021, SEV-2016-0588, SEV-2016-0597, and MDM-2015-0509, some of which include ERDF funds from the European Union. IFAE is partially funded by the CERCA program of the Generalitat de Catalunya.
Research leading to these results has received funding from the European Research
Council under the European Union's Seventh Framework Program (FP7/2007-2013) including ERC grant agreements 240672, 291329, and 306478.
We  acknowledge support from the Brazilian Instituto Nacional de Ci\^encia
e Tecnologia (INCT) do e-Universo (CNPq grant 465376/2014-2).

This manuscript has been authored by Fermi Research Alliance, LLC under Contract No. DE-AC02-07CH11359 with the U.S. Department of Energy, Office of Science, Office of High Energy Physics.

\newcommand{\newblock}{}
\bibliographystyle{jphysicsB.bst}
\bibliography{bib}

\onecolumn{}
\cleardoublepage{}
\pagestyle{fancy}
\renewcommand{\headrulewidth}{0pt}
\rhead{16}
\lhead{\textit{AFFILIATIONS}}
\fancyfoot[]{}
\section*{AFFILIATIONS}
\address{
$^{1}$ Department of Physics \& Astronomy, University College London, Gower Street, London, WC1E 6BT, UK \\
$^{2}$ Department of Astronomy, University of Michigan, Ann Arbor, MI 48109, USA \\
$^{3}$ Department of Physics, University of Michigan, Ann Arbor, MI 48109, USA \\
$^{4}$ Physics Department, 2320 Chamberlin Hall, University of Wisconsin-Madison, 1150 University Avenue Madison, WI  53706-1390 \\
$^{5}$ Cerro Tololo Inter-American Observatory, NSF's National Optical-Infrared Astronomy Research Laboratory, Casilla 603, La Serena, Chile \\
$^{6}$ Departamento de F\'isica Matem\'atica, Instituto de F\'isica, Universidade de S\~ao Paulo, CP 66318, S\~ao Paulo, SP, 05314-970, Brazil \\
$^{7}$ Laborat\'orio Interinstitucional de e-Astronomia - LIneA, Rua Gal. Jos\'e Cristino 77, Rio de Janeiro, RJ - 20921-400, Brazil \\
$^{8}$ Fermi National Accelerator Laboratory, P. O. Box 500, Batavia, IL 60510, USA \\
$^{9}$ Instituto de Fisica Teorica UAM/CSIC, Universidad Autonoma de Madrid, 28049 Madrid, Spain \\
$^{10}$ CNRS, UMR 7095, Institut d'Astrophysique de Paris, F-75014, Paris, France \\
$^{11}$ Sorbonne Universit\'es, UPMC Univ Paris 06, UMR 7095, Institut d'Astrophysique de Paris, F-75014, Paris, France \\
$^{12}$ Kavli Institute for Particle Astrophysics \& Cosmology, P. O. Box 2450, Stanford University, Stanford, CA 94305, USA \\
$^{13}$ SLAC National Accelerator Laboratory, Menlo Park, CA 94025, USA \\
$^{14}$ Instituto de Astrofisica de Canarias, E-38205 La Laguna, Tenerife, Spain \\
$^{15}$ Universidad de La Laguna, Dpto. Astrofísica, E-38206 La Laguna, Tenerife, Spain \\
$^{16}$ Department of Astronomy, University of Illinois at Urbana-Champaign, 1002 W. Green Street, Urbana, IL 61801, USA \\
$^{17}$ National Center for Supercomputing Applications, 1205 West Clark St., Urbana, IL 61801, USA \\
$^{18}$ Institut de F\'{\i}sica d'Altes Energies (IFAE), The Barcelona Institute of Science and Technology, Campus UAB, 08193 Bellaterra (Barcelona) Spain \\
$^{19}$ Jodrell Bank Center for Astrophysics, School of Physics and Astronomy, University of Manchester, Oxford Road, Manchester, M13 9PL, UK \\
$^{20}$ University of Nottingham, School of Physics and Astronomy, Nottingham NG7 2RD, UK \\
$^{21}$ INAF-Osservatorio Astronomico di Trieste, via G. B. Tiepolo 11, I-34143 Trieste, Italy \\
$^{22}$ Institute for Fundamental Physics of the Universe, Via Beirut 2, 34014 Trieste, Italy \\
$^{23}$ Observat\'orio Nacional, Rua Gal. Jos\'e Cristino 77, Rio de Janeiro, RJ - 20921-400, Brazil \\
$^{24}$ Centro de Investigaciones Energ\'eticas, Medioambientales y Tecnol\'ogicas (CIEMAT), Madrid, Spain \\
$^{25}$ Department of Physics, IIT Hyderabad, Kandi, Telangana 502285, India \\
$^{26}$ Santa Cruz Institute for Particle Physics, Santa Cruz, CA 95064, USA \\
$^{27}$ Institute of Theoretical Astrophysics, University of Oslo. P.O. Box 1029 Blindern, NO-0315 Oslo, Norway \\
$^{28}$ Kavli Institute for Cosmological Physics, University of Chicago, Chicago, IL 60637, USA \\
$^{29}$ Institut d'Estudis Espacials de Catalunya (IEEC), 08034 Barcelona, Spain \\
$^{30}$ Institute of Space Sciences (ICE, CSIC),  Campus UAB, Carrer de Can Magrans, s/n,  08193 Barcelona, Spain \\
$^{31}$ Department of Physics, Stanford University, 382 Via Pueblo Mall, Stanford, CA 94305, USA \\
$^{32}$ D\'{e}partement de Physique Th\'{e}orique and Center for Astroparticle Physics, Universit\'{e} de Gen\`{e}ve, 24 quai Ernest Ansermet, CH-1211 Geneva, Switzerland \\
$^{33}$ Department of Physics, ETH Zurich, Wolfgang-Pauli-Strasse 16, CH-8093 Zurich, Switzerland \\
$^{34}$ School of Mathematics and Physics, University of Queensland,  Brisbane, QLD 4072, Australia \\
$^{35}$ Center for Cosmology and Astro-Particle Physics, The Ohio State University, Columbus, OH 43210, USA \\
$^{36}$ Department of Physics, The Ohio State University, Columbus, OH 43210, USA \\
$^{37}$ Faculty of Physics, Ludwig-Maximilians-Universit\"at, Scheinerstr. 1, 81679 Munich, Germany \\
$^{38}$ Max Planck Institute for Extraterrestrial Physics, Giessenbachstrasse, 85748 Garching, Germany \\
$^{39}$ Universit\"ats-Sternwarte, Fakult\"at f\"ur Physik, Ludwig-Maximilians Universit\"at M\"unchen, Scheinerstr. 1, 81679 M\"unchen, Germany \\
$^{40}$ Center for Astrophysics $\vert$ Harvard \& Smithsonian, 60 Garden Street, Cambridge, MA 02138, USA \\
$^{41}$ Australian Astronomical Optics, Macquarie University, North Ryde, NSW 2113, Australia \\
$^{42}$ Lowell Observatory, 1400 Mars Hill Rd, Flagstaff, AZ 86001, USA \\
$^{43}$ George P. and Cynthia Woods Mitchell Institute for Fundamental Physics and Astronomy, and Department of Physics and Astronomy, Texas A\&M University, College Station, TX 77843,  USA \\
$^{44}$ Department of Astrophysical Sciences, Princeton University, Peyton Hall, Princeton, NJ 08544, USA \\
$^{45}$ Instituci\'o Catalana de Recerca i Estudis Avan\c{c}ats, E-08010 Barcelona, Spain \\
$^{46}$ Institute of Astronomy, University of Cambridge, Madingley Road, Cambridge CB3 0HA, UK \\
$^{47}$ Department of Physics and Astronomy, Pevensey Building, University of Sussex, Brighton, BN1 9QH, UK \\
$^{48}$ Department of Physics and Astronomy, University of Pennsylvania, Philadelphia, PA 19104, USA \\
$^{49}$ School of Physics and Astronomy, University of Southampton,  Southampton, SO17 1BJ, UK \\
$^{50}$ Computer Science and Mathematics Division, Oak Ridge National Laboratory, Oak Ridge, TN 37831
}

\end{document}